# Two-dimensional flat-bands in Moiré-diamonds

Yalan Wei[1], Shifang Li[2,#], Yuke Song[1] and Chaoyu He[1,2,*]

[1] School of Physics and Optoelectronics, Xiangtan University, Xiangtan 411105, China

[2] Center for Quantum Science and Technology, Shanghai University, Shanghai 200444, China

Email of Corresponding Author: shifang@shu.edu.cn; hechaoyu@xtu.edu.cn;

The discovery of flat-bands in magic-angle twisted bilayer graphene has underscored the potential of moiré engineering for correlated states, but such phases are notoriously difficult to realize and highly fragile against perturbations. Here, we propose an alternative route to flat-bands by introducing sp³ hybridization in twisted graphite. Instead of relying on fine-tuned magic angles, our approach identifies flat-band states at relatively large twist angles with short moiré periods. In this regime, sp³-induced reconstructions generate electronic states that, once formed, are locked by substantial energy barriers, rendering them robust against external perturbations. Using twisted graphite as a prototype, we uncover a series moiré-diamond that host two-dimensional flat conduction of valence bands, where carriers are localized within specific momentum planes but remain dispersive along orthogonal directions. The emergence of dimensional flat-bands opens a new platform for flat-band-driven correlated physics and suggests opportunities for designing quantum materials with highly directional electronic functionalities.

**Introduction**

The discovery of correlated insulating states and superconductivity in magic-angle twisted bilayer graphene (TBG) [1-3] has opened a new era of moiré physics [4-6], stimulating a surge of research into (2D) materials. Beyond graphene, twisted systems based on hexagonal boron nitride (BN) [7], transition-metal dichalcogenides such as $MoS_2$ [8], $MoTe_2$ [9], $WSe_2$ [10,11], and black phosphorus[12] have been widely explored, demonstrating the versatility of moiré engineering for controlling electronic structures and emergent quantum phases[13-16]. These studies have established flat-band physics as a powerful concept with broad implications for condensed matter physics and materials science.

Nevertheless, moiré flat-bands typically appear only at very small twist angles, where the moiré period becomes extremely large [1-3,17]. This not only renders theoretical calculations and experimental control challenging but also leaves the resulting structures fragile, since their weak van der Waals interlayer interactions are highly susceptible to perturbations. Yet, three-dimensional graphite, van der Waals coupling is not the only possible route to stabilize a specific twist angle. Interlayer hybridization, which converts carbon atoms from $sp^2$ to $sp^3$ coordination, offers stronger interactions to hold twisted stacking. Indeed, experiments on cold-compressed graphite have demonstrated the formation of diamond [18], hexagonal diamond [19-21], and various diamond-like allotropes [22], thereby greatly enriching the structural and property space of carbon [23-26]. Extending this idea to twisted three-dimensional systems [27] raises intriguing questions: can appropriate interlayer hybridization lock in certain twist angles? Where would such phases reside on the potential energy landscape, and what electronic and mechanical properties would they exhibit? Could large-angle twists with short moiré periods generate flat-bands, in analogy to magic-angle physics but within a more robust framework? Addressing these questions is both necessary and timely, and clearly warrants systematic investigation.

In this work, we propose a systematic approach to generate flat-band carbon allotropes by introducing interlayer hybridization into twisted graphite. Based on graph-coloring theory, we assigned alternating up-and-down displacements to carbon atoms in a series of twisted configurations, thereby enforcing $sp^3$ hybridization across layers and forming diamond-like networks. This construction yields a rich family of fully 4-coordinated carbon structures. First-principles calculations reveal that most of these phases are metastable, lying 300–750 meV per atom above the 3C diamond, yet they exhibit an intriguing feature: nearly flat conduction or valence bands confined within the two-dimensional Brillouin zone. Unlike conventional flat-bands, these "2D flat-bands" are flat only in-plane but dispersive along the perpendicular direction, leading to striking anisotropy in carrier velocities. We further investigated several low-energy candidates and confirmed their dynamical and elastic stability, as well as the characteristic real-space patterns of their flat-band states. These results break the conventional constraint that flat-bands in graphene-based systems occur only at small twist angles, while also providing a new strategy for designing flat-band materials through controlled interlayer hybridization. The emergent anisotropic flat-bands may enable novel functionalities in direction-selective transport, quasi-2D correlated electron systems, and anisotropic optoelectronic applications.

**Models and Methods**

Twisted graphite models are constructed by stacking identical hexagonal supercells of graphene monolayer in different orientations, which can be labeled by integer indices (m, n). The hexagonal $\sqrt{N} \times \sqrt{N}$ supercells are governed by indices (m, n) through the relationship[28]:

$$\sqrt{N} = \sqrt{m^2 + n^2 - mn}$$

And the corresponding lattice vectors are given by: $\vec{A} = m\vec{a} + n\vec{b}, \vec{B} = -n\vec{a} + (m-n)\vec{b}$, where $\vec{a} = a(\sqrt{3}/2, -1/2)$ and $\vec{b} = a(0, 1)$ are the lattice vectors for the primitive cell (a=2.46 Å). The corresponding supercell can be generated by applying an expansion matrix $\begin{vmatrix} m & n & 0 \\ -n & (m-n) & 0 \\ 0 & 0 & 0 \end{vmatrix}$ to the primitive cell. It is easy to know that there may exist multiple distinct (m, n) pairs for a given N, for example, N=19 corresponding to both (-5, -3) and (-5, -2). When two different supercells $(m_1, n_1)$ and $(m_2, n_2)$ are stacked together, an effective moiré superlattice can be formed if the twist angle θ is not an integer multiple of 30°. Such a twist angle θ can be calculated by:

$$\theta = \cos^{-1} \frac{m_1 m_2 + n_1 n_2 - \frac{1}{2}(m_1 n_2 + m_2 n_1)}{\sqrt{(m_1^2 + n_1^2 - m_1 n_1)(m_2^2 + n_2^2 - m_2 n_2)}}$$

In our work, we only consider supercells with N<50 in order to keep the total number of atoms below 200. All possible moiré super-cells (N=7, 13, 19, 21, 28, 31, 37, 39, 43, and 49) and their effective stacking combinations are listed in Supplementary Table S1.

We then construct interlayer hybridized structures based on a random coloring strategy. As shown in Fig. S1, each moiré superlattice is first loaded into the RG2[29-31] and associated with the in-plane 3-coordinated quotient graph. RG2 then determines the symmetry and subgroups of each structure. On this basis, random up-and-down colorings are applied to the inequivalent atoms to generate buckled configurations. After forming the buckled configurations, interlayer atomic connections are established following the nearest-neighbor principle, leading to new 4-coordinated structures. In our search, the smallest carbon rings are constrained to 5-membered to exclude high-energy 3- and 4-membered ones. For perturbed structures that successfully establish a 4-coordinated quotient graph, RG2 performs rapid preliminary optimization to adjust bond lengths and bond angles as close as possible to 1.56 Å and 109°, as in the diamond structure. Structures that satisfy these geometric constraints are retained, deduplicated, and subsequently subjected to further high-precision optimization using first-principles methods.

In total, we obtained 180 Moiré-diamond structures that satisfy the selection criteria and performed structural optimization and stability assessments using first-principles calculations. All simulations were carried out with the VASP package[32], employing the plane-wave pseudopotential method together with the PBE exchange–correlation functional[33,34]. A plane-wave cutoff energy of 500 eV was adopted, and the Brillouin zone was sampled using k-point grids corresponding to a maximum spacing finer than 0.21 1/Å. The convergence criteria were set to $10^{-6}$ eV for the electronic self-consistency loop and 0.01 eV/Å for the residual forces. Dynamical and elastic stability were evaluated within the framework of density functional perturbation theory (DFPT)[35], from which phonon dispersions and elastic constants were obtained. Phonon spectra were calculated with the aid of the Phonopy package[36]. For high-throughput electronic structure calculations, we further employed the general tight-binding method (gt-TB)[37-40], which provides an efficient and accurate approximation to DFT while enabling large-scale screening.

**Results and discussion**

To assess the relative stability of these moiré-diamonds, 3C diamond is taken as a reference, and the average energy (eV/atom) versus average volume (Å³/atom) is shown in Fig. S2 (a). Most structures lie 300–500 meV/atom above 3C diamond, yet remain more stable than the well-known BC8 (695 meV/atom) [41,42] and experimentally synthesized T-carbon (1171 meV/atom) [43,44]. As summarized in table S2 and Fig. S2 (b), these structures are mainly distributed in space groups 143 (21), 150 (16), 158 (42), 159 (63), and 163 (10), with fewer in 147 (9), 149 (1), 165 (6), 168 (8) and 177 (4). Size-wise, they cluster at N = 39 (54), 37 (38), 21 (13), 19 (25), and 13 (19), with only a few at N = 7 (4), 28 (9), 31 (11), 43 (6), and 49 (1). Despite extensive sampling, the distributions exhibit no clear regularity, likely due to the interplay of symmetry and state-space size. Larger supercells expand the state space, lowering the probability of sampling favorable configurations, while higher-symmetry subgroups limit corrugation patterns and interlayer hybridization. Some subgroups (e.g., 192, 190, and 188) do not permit interlayer corrugation at all. Overall, a diverse set of moiré-diamond structures in trigonal and hexagonal systems is obtained. Their CIF files moiré-diamond structures are provided as supplementary data. Eight of the most stable ones, including the previously predicted m-dia (147-6-28-r568-0) are shown in Fig. S3, whose [001] projections clearly display pronounced moiré patterns. Following the established conventions for naming carbon crystal structures[29], we designate these carbon phases based on their space group, number of inequivalent atoms, total number of atoms, ring topology, and duplicate index.

Using the gt-TB software developed by our group[37-39] and well-established parameters for $sp^3$ carbon[40], high-throughput calculations of the 2D band structures (along the conventional high-symmetry lines) and 3D band structures (in the entire 2D first Brillouin zone at Z = 0) for these moiré-diamond structures. The corresponding band structures are shown in Supplementary Figs. S4–S18. Based on the dispersion of the first conduction band ($CB_1$) and the first valence band ($VB_1$) in the 3D band structures, we classified the structures according to a bandwidth criterion of less than 30 meV. Specifically, the four categories include: CBM flat-band structures ($CB_1$ dispersion < 30 meV, $VB_1$ dispersion ≥ 30 meV), VBM flat-band structures ($VB_1$ dispersion < 30 meV, $CB_1$ dispersion ≥ 30 meV), CBM-VBM flat-band structures (both $CB_1$ and $VB_1$ dispersion < 30 meV), and normal semiconductors (both $CB_1$ and $VB_1$ dispersion ≥ 30 meV). The energy and bandgap diagrams are shown in Fig. 1(a), where each type is distinguished by different symbols. Finally, we identified 30 CBM flat-band structures, 27 VBM flat-band structures, and 26 CBM-VBM flat-band structures. As summarized in Fig. 1(b), two-dimensional flat-bands appear at either the CBM or VBM for all sizes, except for the two smallest sizes, N=7 and N=13. The results in Fig. 1(c) suggest that their distribution seems to be independent of symmetry. These results suggest that, unlike magic-angle graphene, where flat bands form only under specific small-angle superlattices, moiré-diamonds with interlayer hybridization exhibit significantly enhanced interlayer interactions. This allows flat bands to emerge across a broader range of large twist angles, corresponding to short periodicities. Such flexibility in twist angles could offer new opportunities for engineering electronic properties and exploring strongly correlated phenomena. Notably, the 2D flat bands in these 3D materials display highly anisotropic transport behavior, with carriers localized in the two in-plane directions while remaining dispersive along the perpendicular direction. This unique anisotropy could have profound implications for electronic transport and correlated phenomena in these systems.

Please note that we just performed finite-time sampling of moiré-diamond structures within limited system sizes. Although the results are not exhaustive, they reveal clear trends: flat-bands can emerge under multiple twist conditions and symmetries. Among all examined systems, the most stable flat-band state appears at the √37×√37 cell (163-13-148-r6-0), which also hosts the most representative structure identified in our search. Notably, this phase has the lowest energy within this size and ranks fourth-lowest across all sizes, making it particularly intriguing for further investigation. As shown in Fig. 2, the optimized crystalline structures of 163-13-148-r6-0 are presented in both top and side views, with the corresponding phonon spectrum displayed below. In the figure, the two graphite layers, marked by red and blue spheres, exhibit a pronounced moiré pattern in their top views. Notably, after interlayer hybridization, the entire structure remains composed exclusively of six-membered rings—a relatively rare geometric feature. This all-six-membered-ring configuration can be extended to other binary systems, such as BN and SiC, offering intriguing possibilities for further exploration. As a potential new carbon phase, the formed moiré-diamond (163-13-148-r6-0) is dynamically stable, as evidenced by its phonon spectrum, which shows no imaginary frequencies. Similarly, it is mechanically stable, as its elastic constants (GPa) $C_{11} = 1097, C_{12} = 109, C_{13} = 83, C_{14} = 0, C_{15} = 5.22, C_{33} = 1098$ $and$ $C_{44} = 467$ satisfy the Born–Huang criteria[45].

To illustrate the electronic properties of structure 163-13-148-r6-0, Fig. 3 (a) presents its 2D band structure, calculated using gt-TB along the conventional high-symmetry lines, along with the 3D band structures of $VB_1$ and $CB_1$ across the 2D first Brillouin zone. The results reveal that this material is an indirect semiconductor with a band gap of 3.659 eV. Both the valence band maximum (VBM) and conduction band minimum (CBM) are located within the kz=0 plane of the 2D Brillouin zone. It is clearly that $CB_1$ in the 2D band structure exhibits a distinct flat-band segment with an energy fluctuation of approximately 9 meV. And this can be further confirmed by the corresponding 3D band structure of $CB_1$ across the entire 2D Brillouin zone. This small fluctuation clearly meets the criteria for a flat-band, indicating that $CB_1$ is a flat-band. In contrast, $VB_1$ shows significant fluctuations in the 2D band structure, with the energy variation across the entire 2D Brillouin zone reaching 318 meV, thus disqualifying it as a flat-band.

To validate this unusual electronic structure, we also calculated the PBE-based band structure of 163-13-148-r6-0 as the results shown in Fig. 3 (b). The results show that $CB_1$ remains nearly flat across the entire 2D Brillouin zone, with an energy variation of about 6 meV, consistent with the TB calculation (9 meV). The main difference is that the PBE-based band gap, 2.724 eV, is significantly smaller than 3.659 eV calculated from TB. This discrepancy arises because the TB parameters were fitted to HSE calculations, whose accuracy has been validated in previous work [40]. Moreover, the difference between the PBE and TB band gaps, 1.135 eV, is consistent with values reported in most previous studies for carbon [23-26].

Notably, this flat-band differs from most previously reported cases: it is not completely flat across the 3D Brillouin zone, but only within the 2D Brillouin zone, exhibiting dispersion along the z direction. This indicates that carriers at the conduction band minimum are localized in the x-y plane and only mobile along z, implying a pronounced anisotropy that can significantly affect transport properties. To elucidate the origin of this flat band, we plotted the corresponding real-space wave function distribution, as shown in Fig. 3(b). The wave functions are primarily localized around the six-fold symmetric centers of the structure (the threefold rotoinversion axes), remaining well-separated in the x-y plane with no interconnections, while extending continuously along the z direction. This indicates that the flat band is confined in the x-y plane but conductive along z, providing a qualitative explanation for the emergence of the 2D flat band.

In summary, we have systematically constructed and screened a large set of moiré-diamond structures derived from twisted hexagonal graphene supercells. High-throughput calculations, based on first-principles methods and the gt-TB approach, reveal that interlayer hybridization can generate robust 2D flat bands in 3D materials across a wide range of twist angles. The most stable flat-band structure, 163-13-148-r6-0, exhibits pronounced anisotropy: carriers are localized within the x–y plane but remain dispersive along z, as confirmed by real-space wave function analysis. This structure is both dynamically and mechanically stable, retaining exclusively six-membered rings after interlayer hybridization. These results establish moiré diamonds as a versatile platform for engineering flat-band physics beyond magic-angle graphene, offering new opportunities to explore anisotropic transport and strongly correlated phenomena in three-dimensional carbon systems.

**Acknowledgments**

This work is supported by the National Natural Science Foundation of China (Grant Nos. 52372260, and 12204397, 12374046), the Youth Science and Technology Talent Project of Hunan Province (Grant No. 2022RC1197) and the Research Foundation of Education Bureau of Hunan Province, China (Grant No. 24A0121), the Science Fund for Distinguished Young Scholars of Hunan Province of China (No.2024JJ2048).

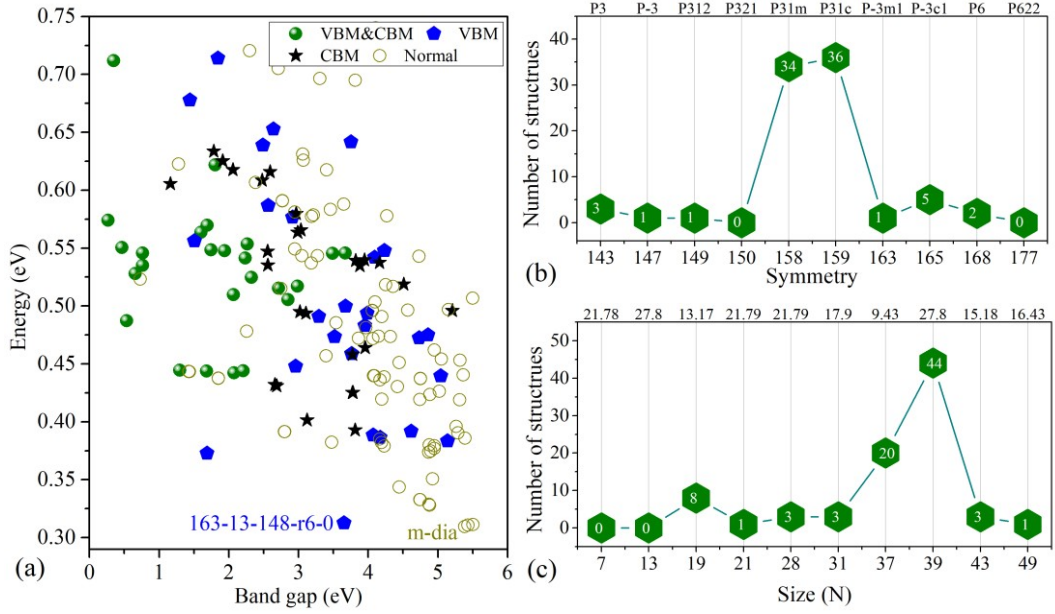

Fig 1. (a) Scatter plots of the DFT-based relative average energies (eV/atom) versus gt-TB based band gaps (eV) of the 180 Moiré-diamonds, where the green solid circles, blue pentagons, black stars, and light yellow circles correspond to VBM-CBM flat-band, VBM flat-band, CBM flat-band semiconductors, and normal semiconductors, respectively. The distributions of flat-band semiconductors with different symmetries and twist supercell sizes are likewise presented in panels (b) and (c).

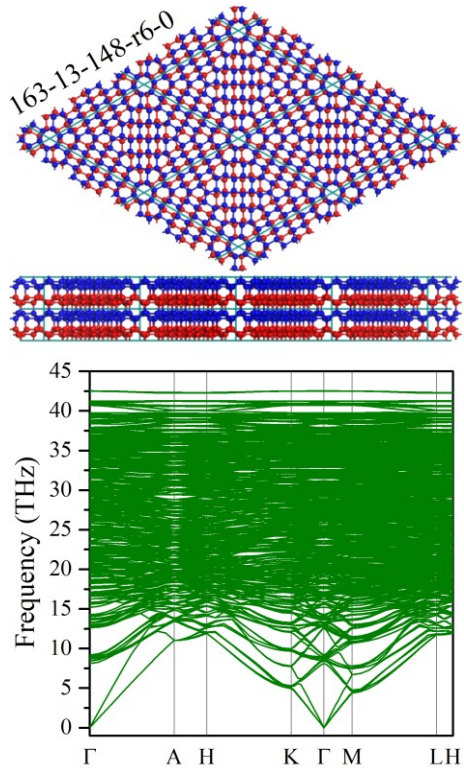

Fig 2. Upper panel: top and side views of 163-13-148-r6-0 along [001] direction. where blue and solid balls are used to distinguish different layers of grpahene. Upper panel: vibrational spectrums for163-13-148-r6-0.

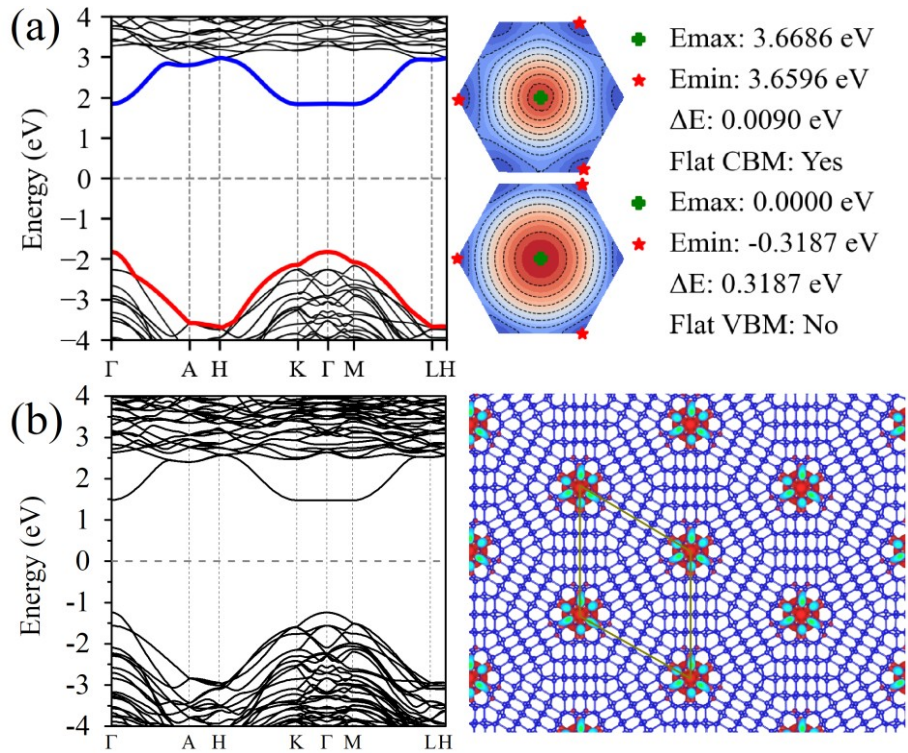

Fig 3. (a) TB-based 2D band structures along conventional high-symmetry paths and 3D representations of the highest valence and lowest conduction bands sampled in the Kz=0 plane of the 2D BZ for 163-13-148-r6-0. (b) PBE-based 2D band structures along conventional high-symmetry paths for 163-13-148-r6-0 and and the real-space partial charge density distribution of the CBM flat-bands.

# SI for "Two-dimensional flat-bands in Moiré-diamond"

Yalan Wei[1], Shifang Li[2,#], Yuke Song[1] and Chaoyu He[1,2,*]

[1] School of Physics and Optoelectronics, Xiangtan University, Xiangtan 411105, China

[2] Center for Quantum Science and Technology, Shanghai University, Shanghai 200444, China

Email of Corresponding Author: shifang@shu.edu.cn; hechaoyu@xtu.edu.cn;

**Table S1** Basic information of various effective twisted supercells in size of $\sqrt{N} \times \sqrt{N}$

| Size (N) / Information | 7 | 13 | 19 | 21 | 28 | 31 | 37 | 39 | 43 | 49 |
|---|---|---|---|---|---|---|---|---|---|---|
| θ | 21.78 | 27.80 | 13.17 | 21.79 | 21.79 | 17.90 | 9.43 | 27.80 | 15.18 | 16.43 |
| (m1,n1) | 3,2 | 4,3 | 5,3 | 5,4 | 6,4 | 6,5 | 7,4 | 7,5 | 7,6 | 8,5 |
| (m2,n2) | 3,1 | 4,1 | 5,2 | 5,1 | 6,2 | 6,1 | 7,3 | 7,2 | 7,1 | 8,3 |
| $N_{atom}$ | 28 | 52 | 76 | 84 | 112 | 124 | 148 | 156 | 172 | 196 |

**Table S2** Statistical table of moiré diamonds with different symmetries obtained from the hybridization of twisted graphite at various supercells of $\sqrt{N} \times \sqrt{N}$

| Size (N) / Symmetry | 7 | 13 | 19 | 21 | 28 | 31 | 37 | 39 | 43 | 49 | Tot |
|---|---|---|---|---|---|---|---|---|---|---|---|
| 143 (P3) | 0 | 8 | 4 | 3 | 0 | 0 | 6 | 0 | 0 | 0 | **21** |
| 147 (P-3) | 1 | 3 | 3 | 2 | 0 | 0 | 0 | 0 | 0 | 0 | **9** |
| 149 (P312) | 0 | 0 | 0 | 0 | 1 | 0 | 0 | 0 | 0 | 0 | **1** |
| 150 (P321) | 1 | 6 | 6 | 2 | 1 | 0 | 0 | 0 | 0 | 0 | **16** |
| 158 (P31m) | 0 | 0 | 0 | 1 | 0 | 0 | 0 | 41 | 0 | 0 | **42** |
| 159 (P31c) | 0 | 1 | 8 | 3 | 3 | 11 | 24 | 8 | 4 | 1 | **63** |
| 163 (P-3m1) | 1 | 1 | 2 | 0 | 2 | 0 | 4 | 0 | 0 | 0 | **10** |
| 165 (P-3c1) | 0 | 0 | 0 | 1 | 0 | 0 | 0 | 5 | 0 | 0 | **6** |
| 168 (P6) | 0 | 0 | 2 | 0 | 0 | 0 | 4 | 0 | 2 | 0 | **8** |
| 177 (P622) | 1 | 0 | 0 | 1 | 2 | 0 | 0 | 0 | 0 | 0 | **4** |
| Tot | **4** | **19** | **25** | **13** | **9** | **11** | **38** | **54** | **6** | **1** | **180** |

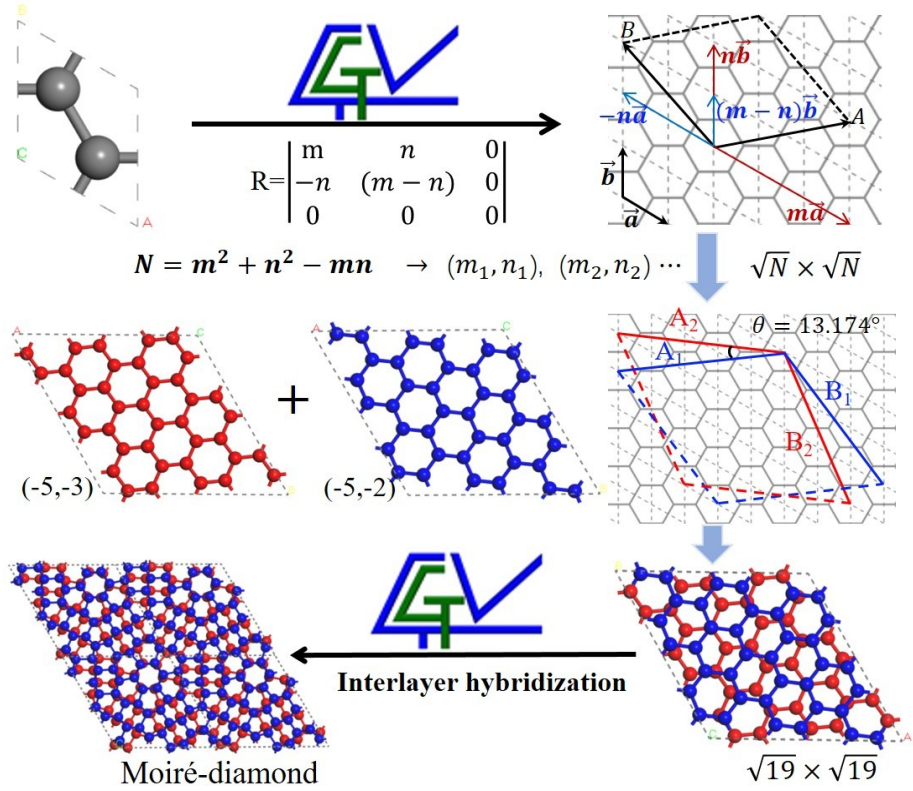

Fig S1. Workflow for sampling moiré diamonds arising from twisted graphite interlayer hybridization with RG2.

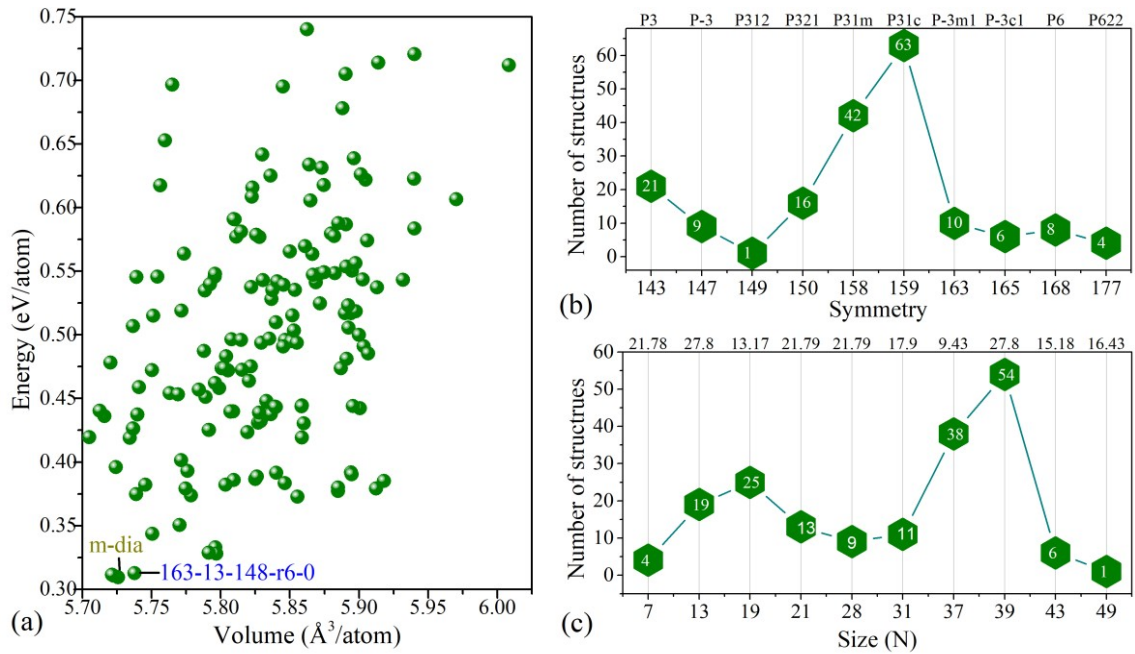

Fig. S2. (a) Scatter plots of the relative average energies (eV/atom) versus equilibrium volumes (Å³/atom) of the 180 Moiré-diamonds calculated by PBE functional. (b) and (c) show the distributions of these structures in different symmetries and under different size (twisted angle) conditions, respectively.

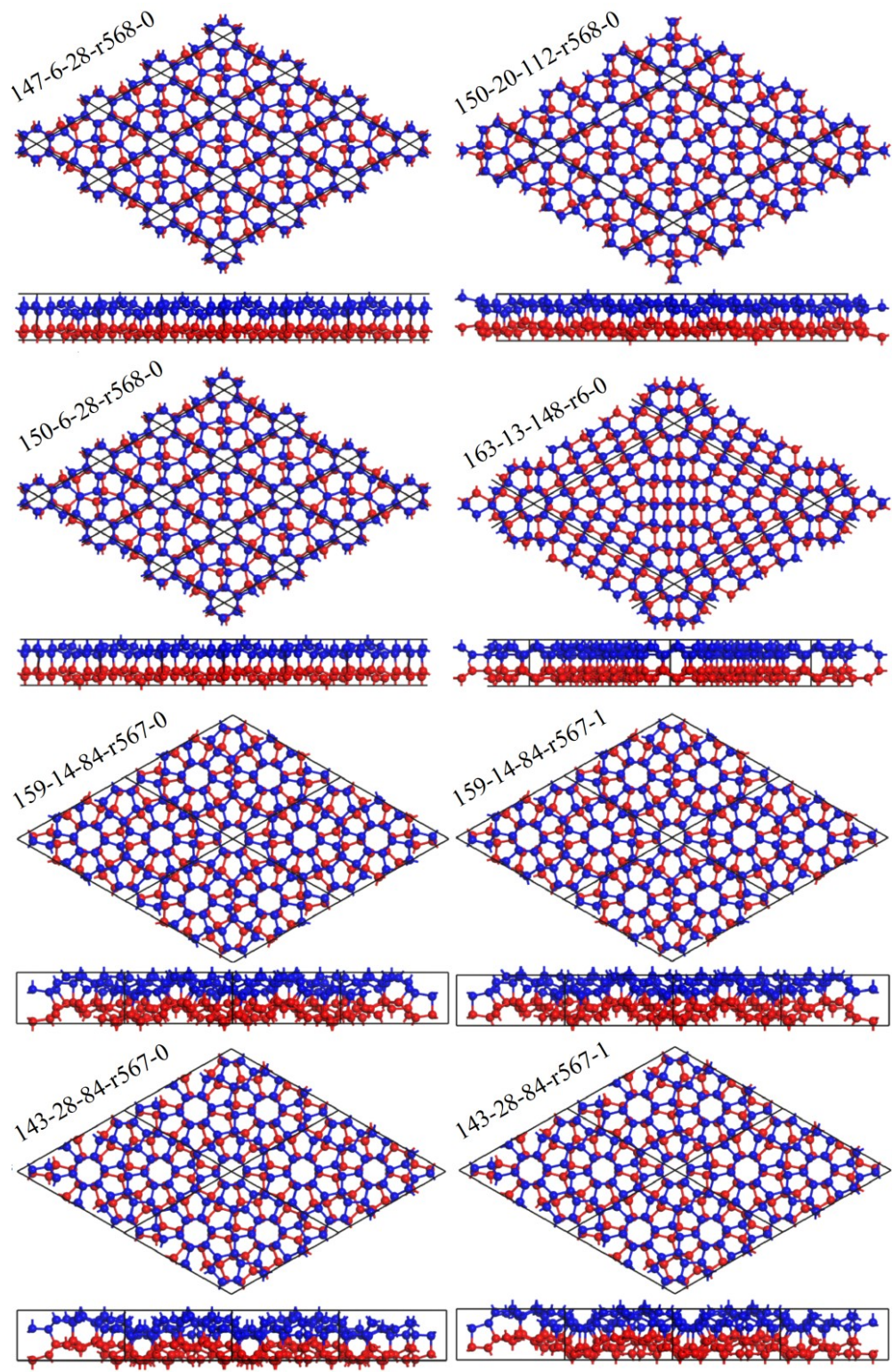

Fig S3. Crystal structures of the eight most stable moiré diamonds sampled in this work are shown in top and side views along the [001] direction. Red and blue balls represent atoms from different graphene layers.

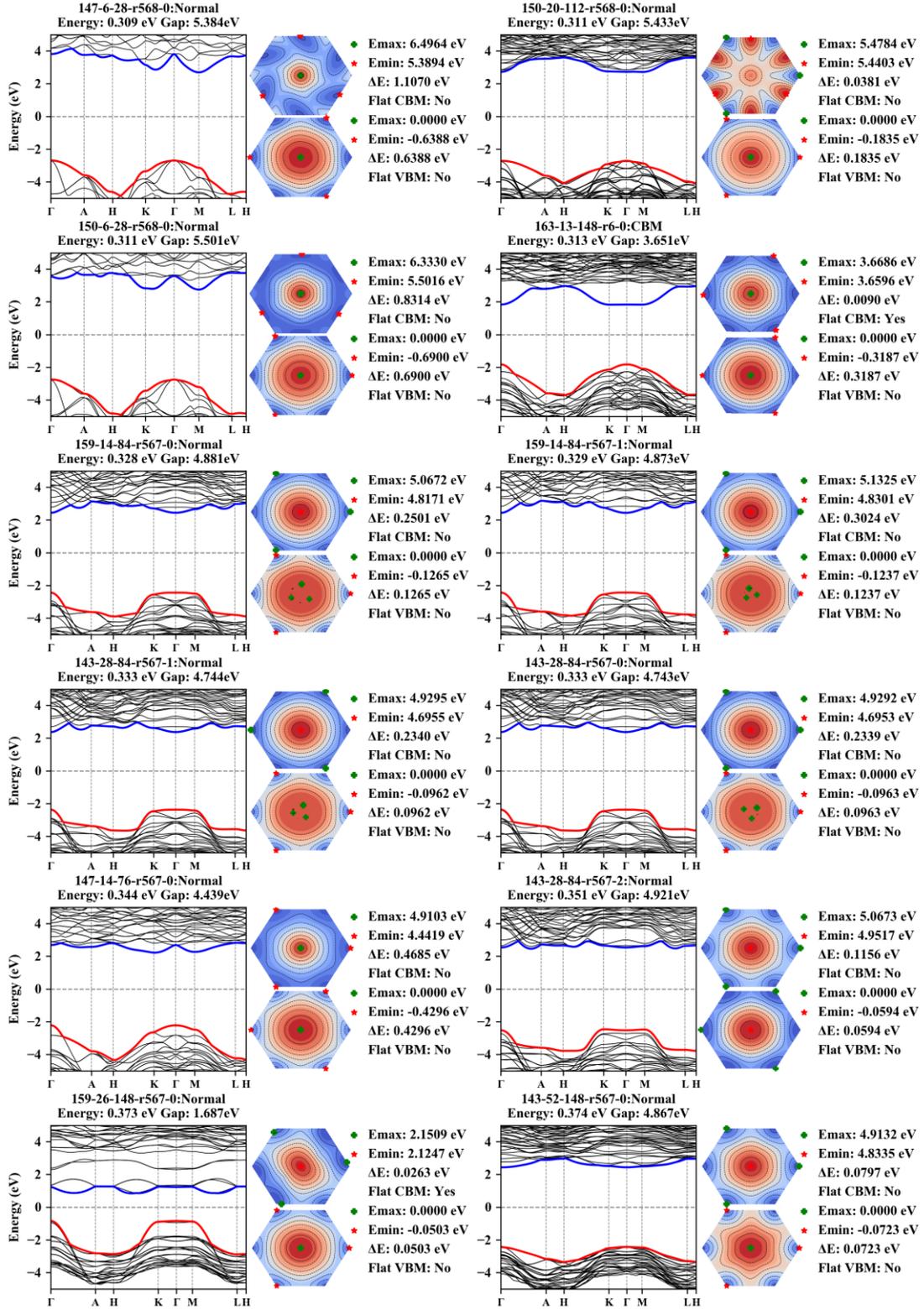

Fig S4. (Structures ranked 1–12 in terms of stability) 2D tight-binding band structures along conventional high-symmetry paths and 3D representations of the highest valence and lowest conduction bands sampled in the Kz=0 plane of the 2D BZ. The total energy, band gap, maxima and minima of valence and conduction bands, band fluctuations in the 2D BZ are provided.

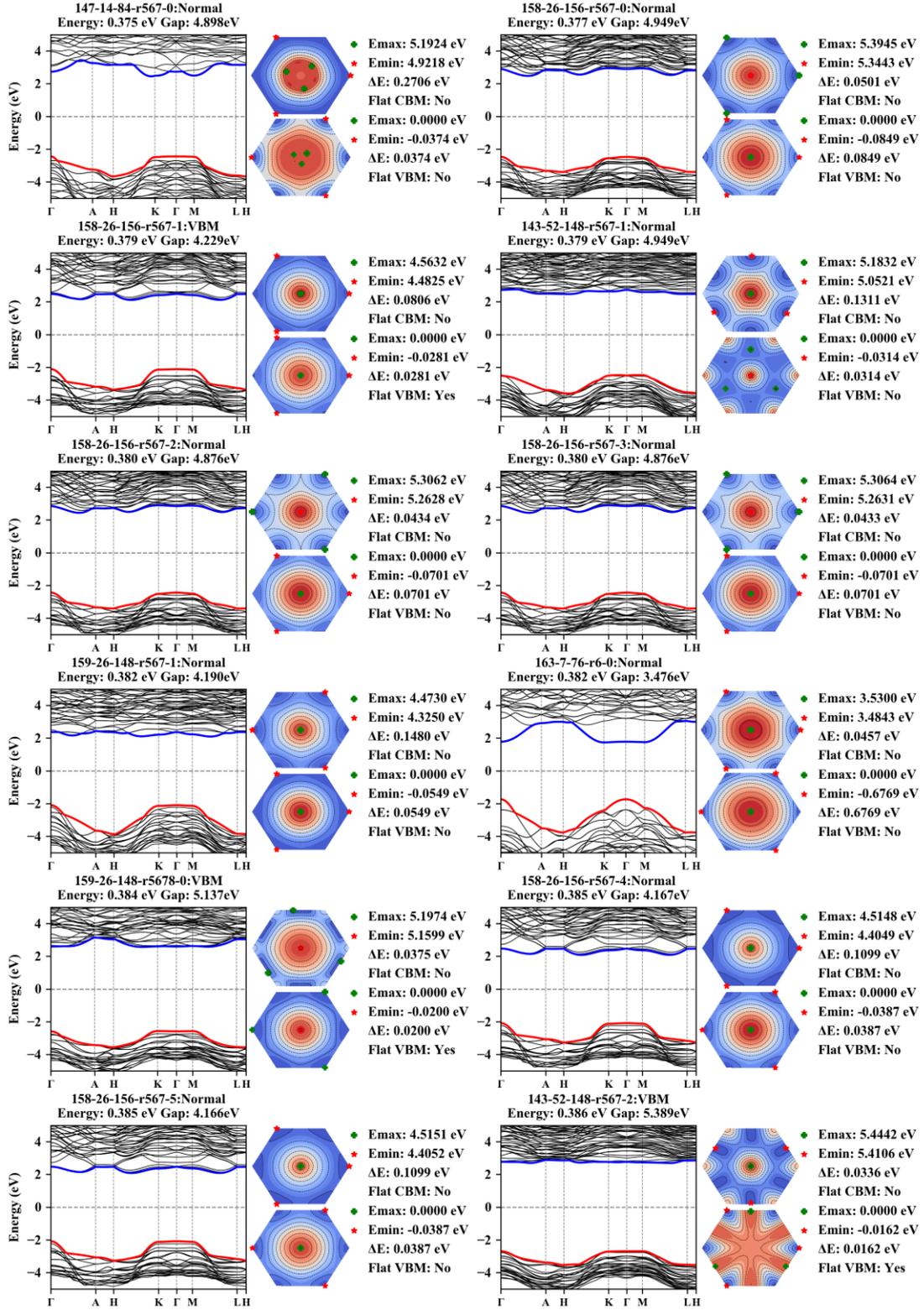

Fig S5. (Structures ranked 13–24 in terms of stability) 2D tight-binding band structures along conventional high-symmetry paths and 3D representations of the highest valence and lowest conduction bands sampled in the Kz=0 plane of the 2D BZ. The total energy, band gap, maxima and minima of valence and conduction bands, band fluctuations in the 2D BZ are provided.

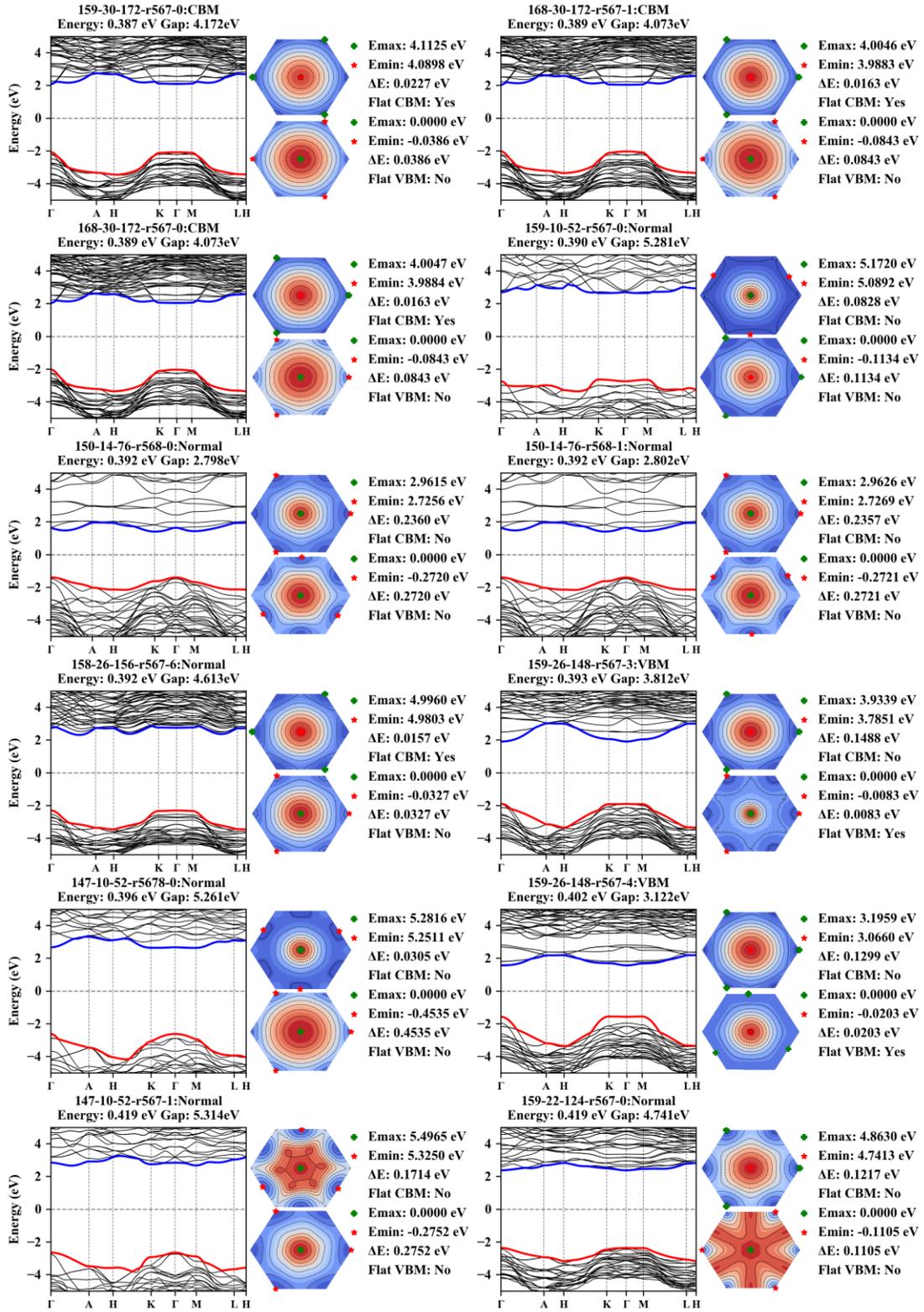

Fig S6. (Structures ranked 25–36 in terms of stability) 2D tight-binding band structures along conventional high-symmetry paths and 3D representations of the highest valence and lowest conduction bands sampled in the Kz=0 plane of the 2D BZ. The total energy, band gap, maxima and minima of valence and conduction bands, band fluctuations in the 2D BZ are provided.

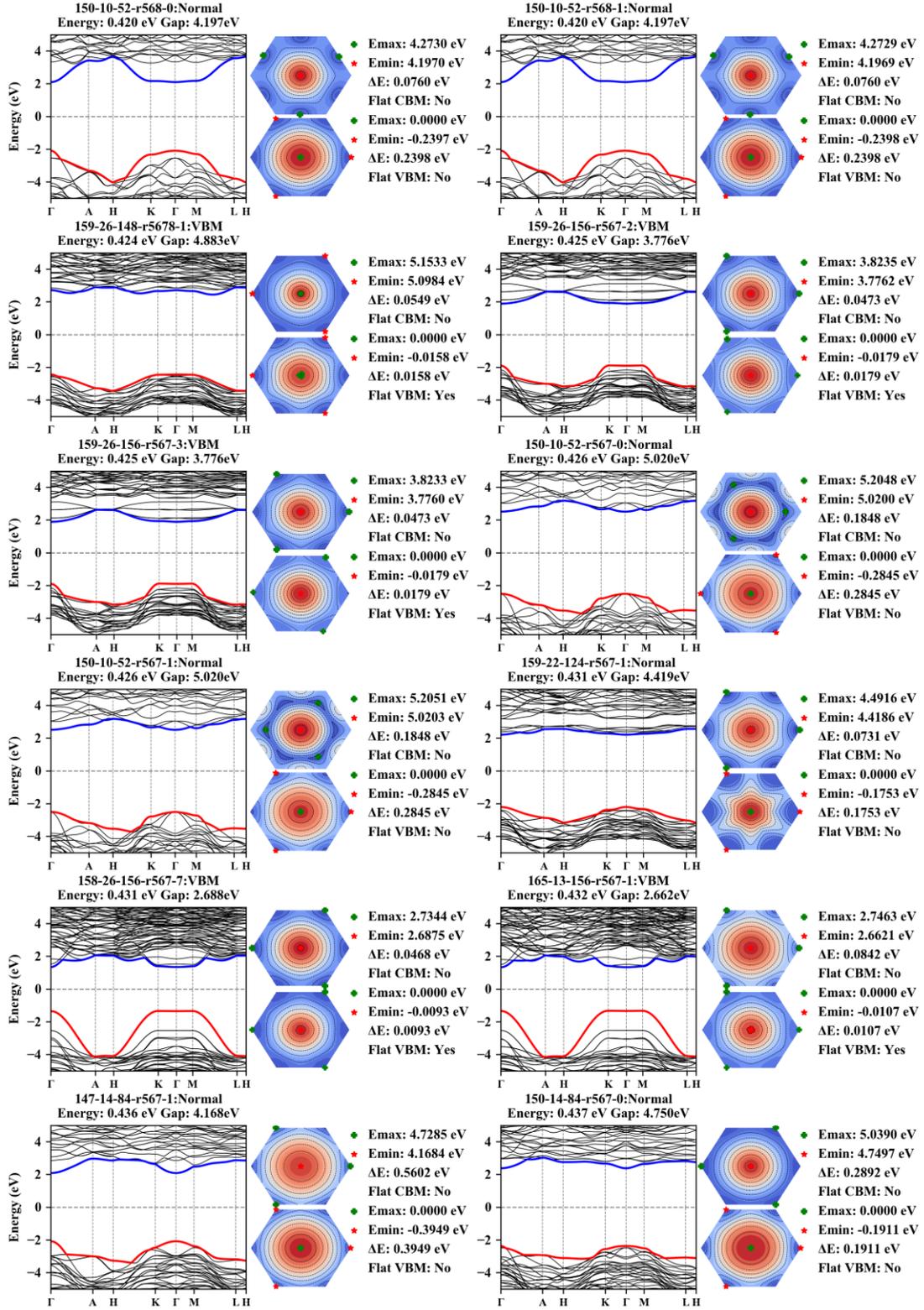

Fig S7. (Structures ranked 37–48 in terms of stability) 2D tight-binding band structures along conventional high-symmetry paths and 3D representations of the highest valence and lowest conduction bands sampled in the Kz=0 plane of the 2D BZ. The total energy, band gap, maxima and minima of valence and conduction bands, band fluctuations in the 2D BZ are provided.

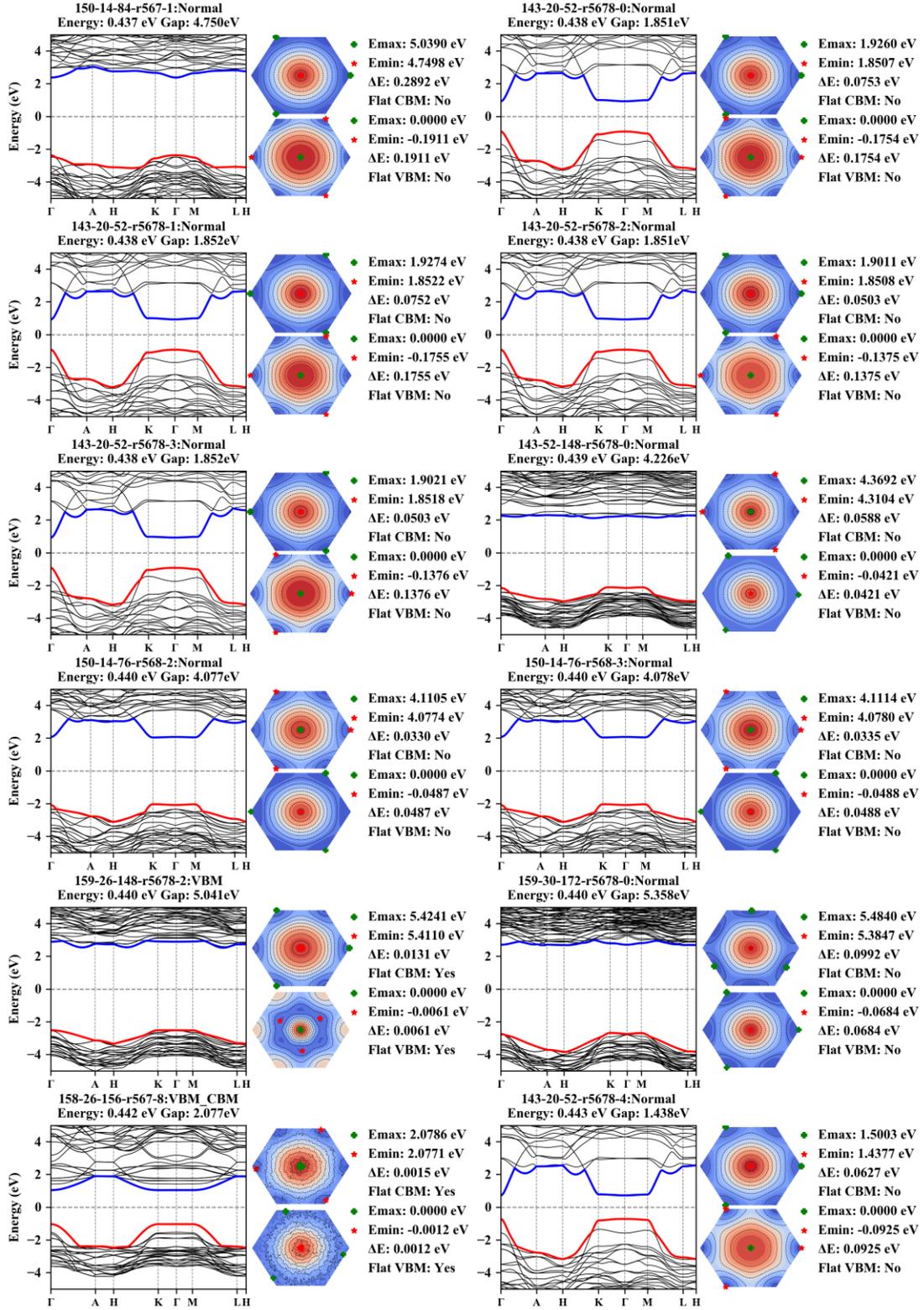

Fig S8. (Structures ranked 49–60 in terms of stability) 2D tight-binding band structures along conventional high-symmetry paths and 3D representations of the highest valence and lowest conduction bands sampled in the Kz=0 plane of the 2D BZ. The total energy, band gap, maxima and minima of valence and conduction bands, band fluctuations in the 2D BZ are provided.

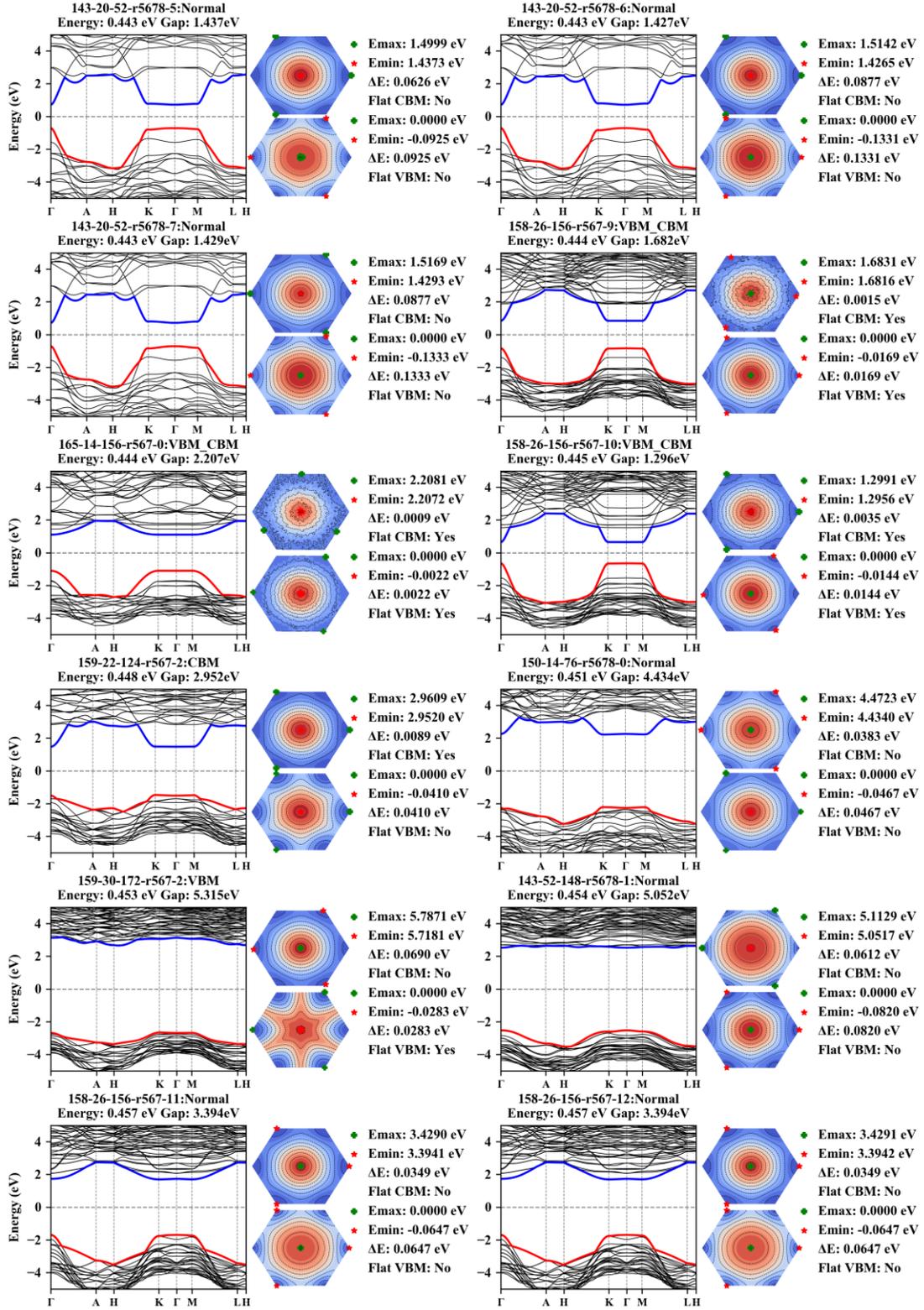

Fig S9. (Structures ranked 61–72 in terms of stability) 2D tight-binding band structures along conventional high-symmetry paths and 3D representations of the highest valence and lowest conduction bands sampled in the Kz=0 plane of the 2D BZ. The total energy, band gap, maxima and minima of valence and conduction bands, band fluctuations in the 2D BZ are provided.

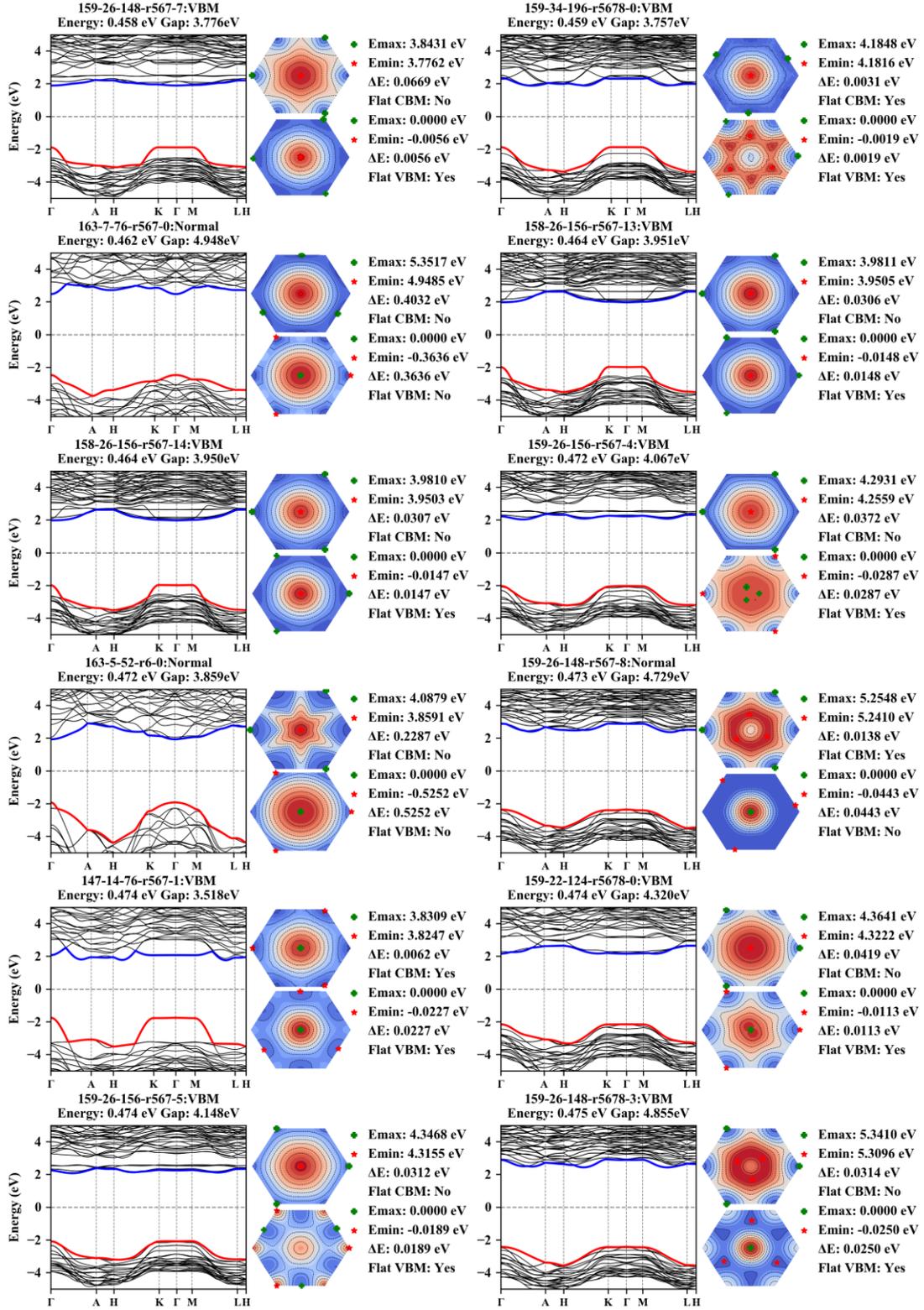

Fig S10. (Structures ranked 73–84 in terms of stability) 2D tight-binding band structures along conventional high-symmetry paths and 3D representations of the highest valence and lowest conduction bands sampled in the Kz=0 plane of the 2D BZ. The total energy, band gap, maxima and minima of valence and conduction bands, band fluctuations in the 2D BZ are provided.

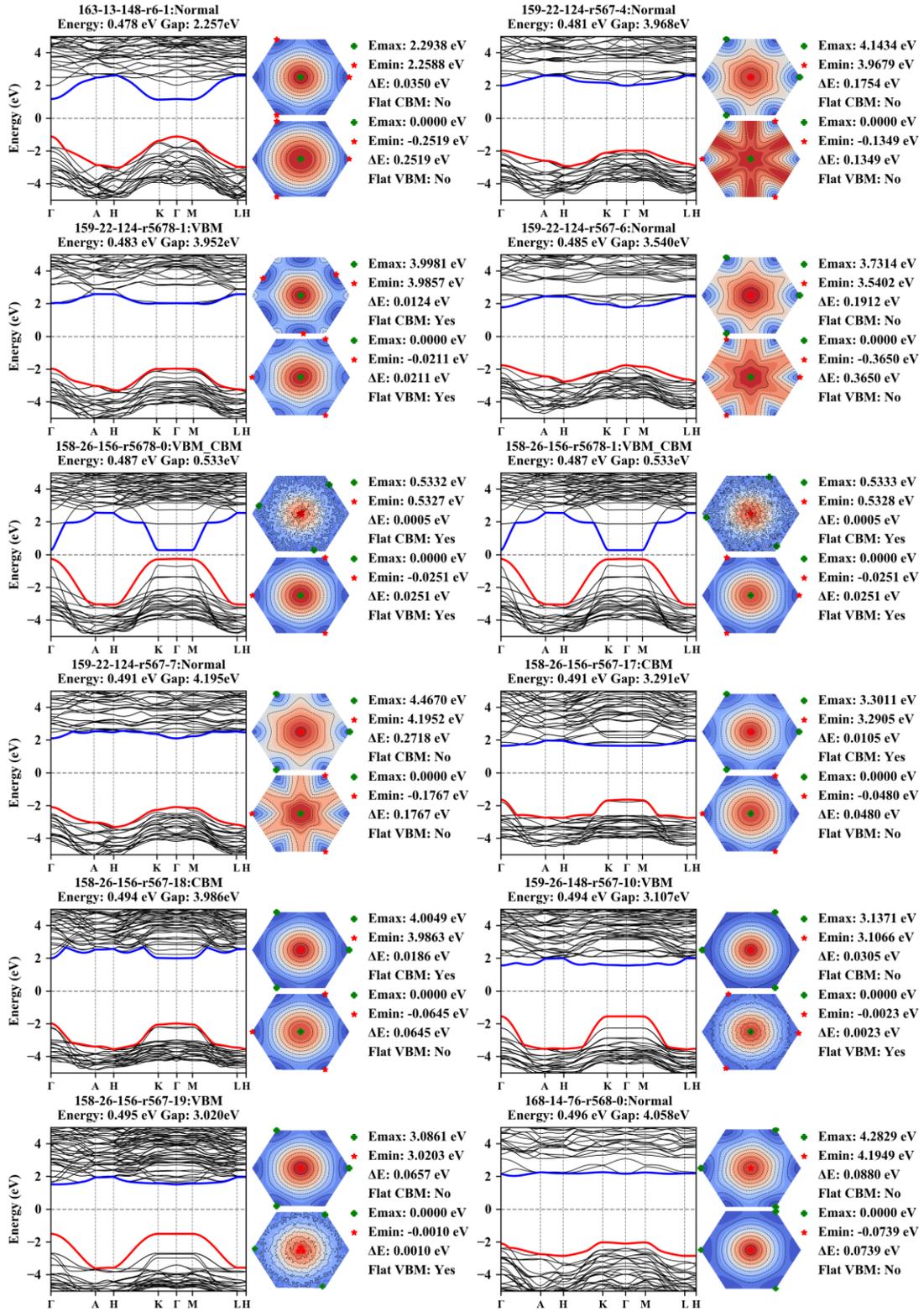

Fig S11. (Structures ranked 85–96 in terms of stability) 2D tight-binding band structures along conventional high-symmetry paths and 3D representations of the highest valence and lowest conduction bands sampled in the Kz=0 plane of the 2D BZ. The total energy, band gap, maxima and minima of valence and conduction bands, band fluctuations in the 2D BZ are provided.

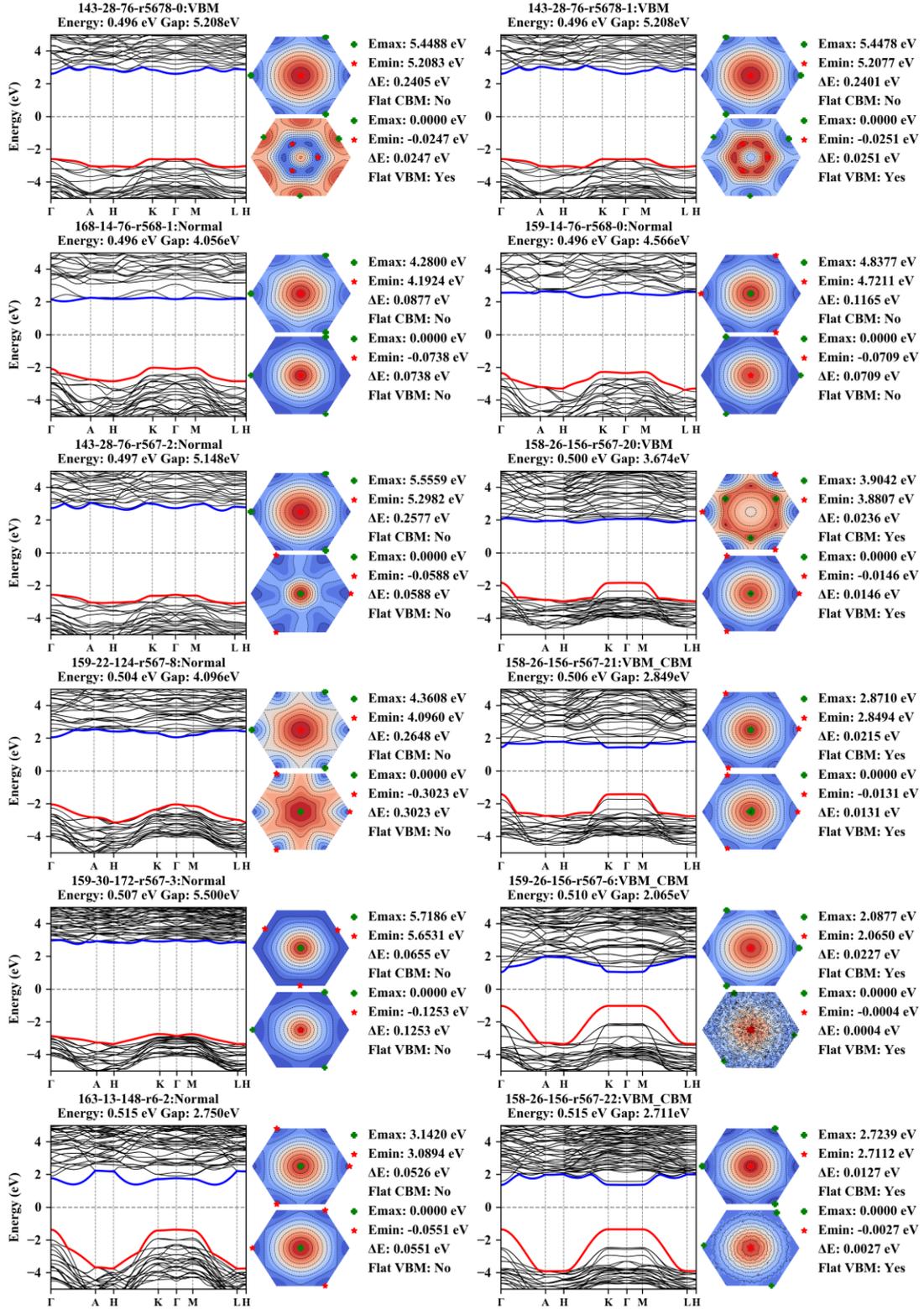

Fig S12. (Structures ranked 97–108 in terms of stability) 2D tight-binding band structures along conventional high-symmetry paths and 3D representations of the highest valence and lowest conduction bands sampled in the Kz=0 plane of the 2D BZ. The total energy, band gap, maxima and minima of valence and conduction bands, band fluctuations in the 2D BZ are provided.

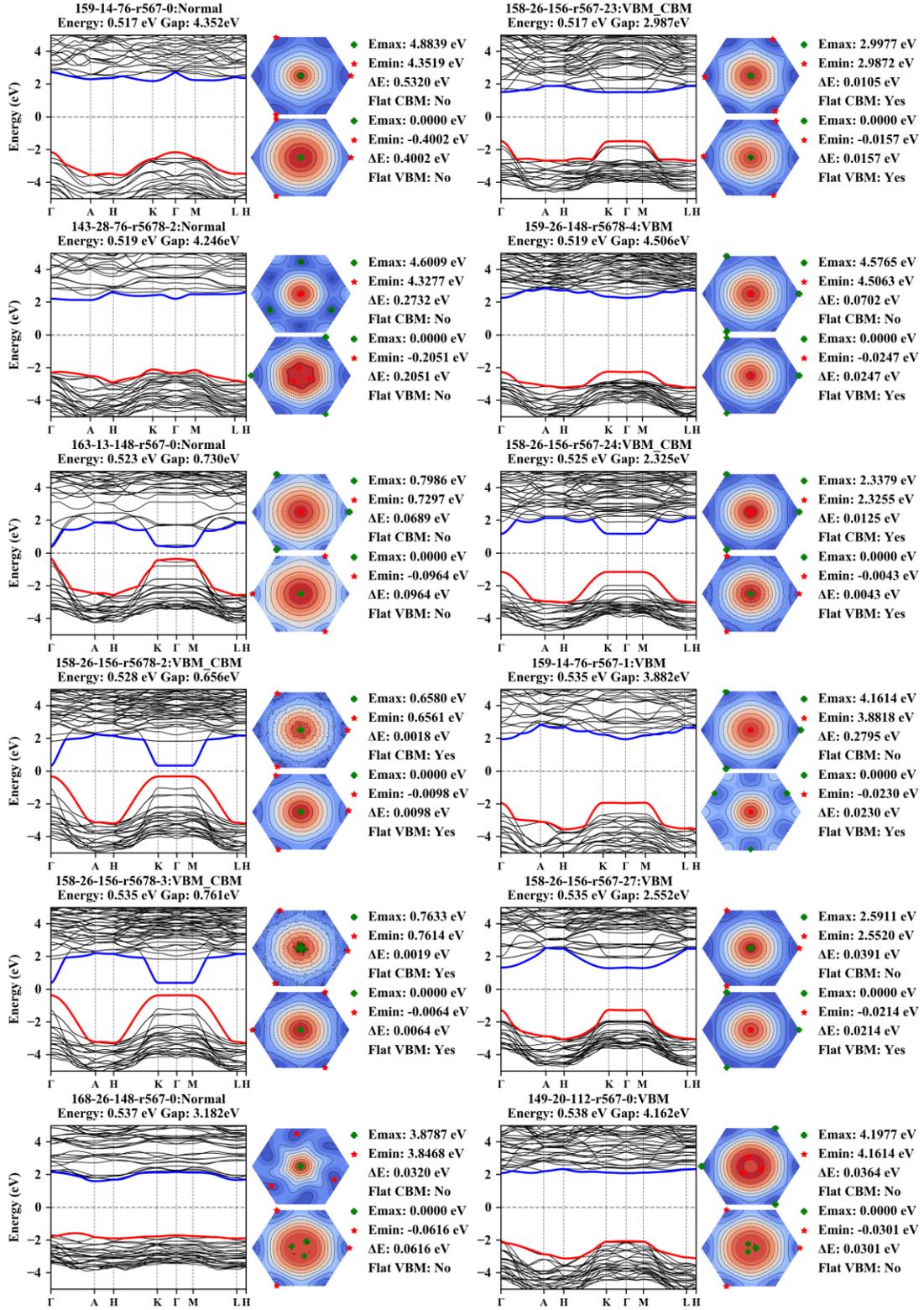

Fig S13. (Structures ranked 109–120 in terms of stability) 2D tight-binding band structures along conventional high-symmetry paths and 3D representations of the highest valence and lowest conduction bands sampled in the Kz=0 plane of the 2D BZ. The total energy, band gap, maxima and minima of valence and conduction bands, band fluctuations in the 2D BZ are provided.

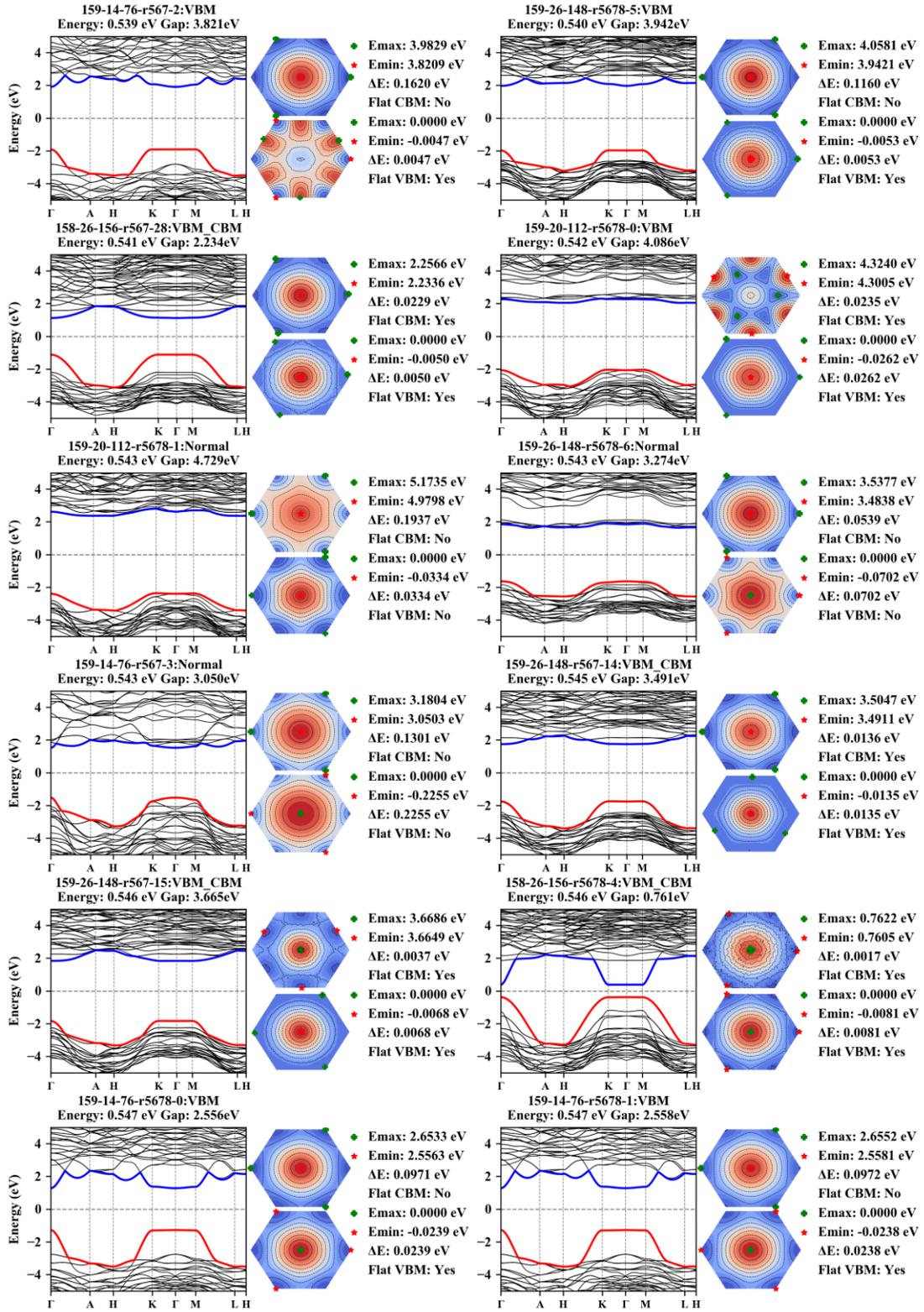

Fig S14 (Structures ranked 121–132 in terms of stability) 2D tight-binding band structures along conventional high-symmetry paths and 3D representations of the highest valence and lowest conduction bands sampled in the Kz=0 plane of the 2D BZ. The total energy, band gap, maxima and minima of valence and conduction bands, band fluctuations in the 2D BZ are provided.

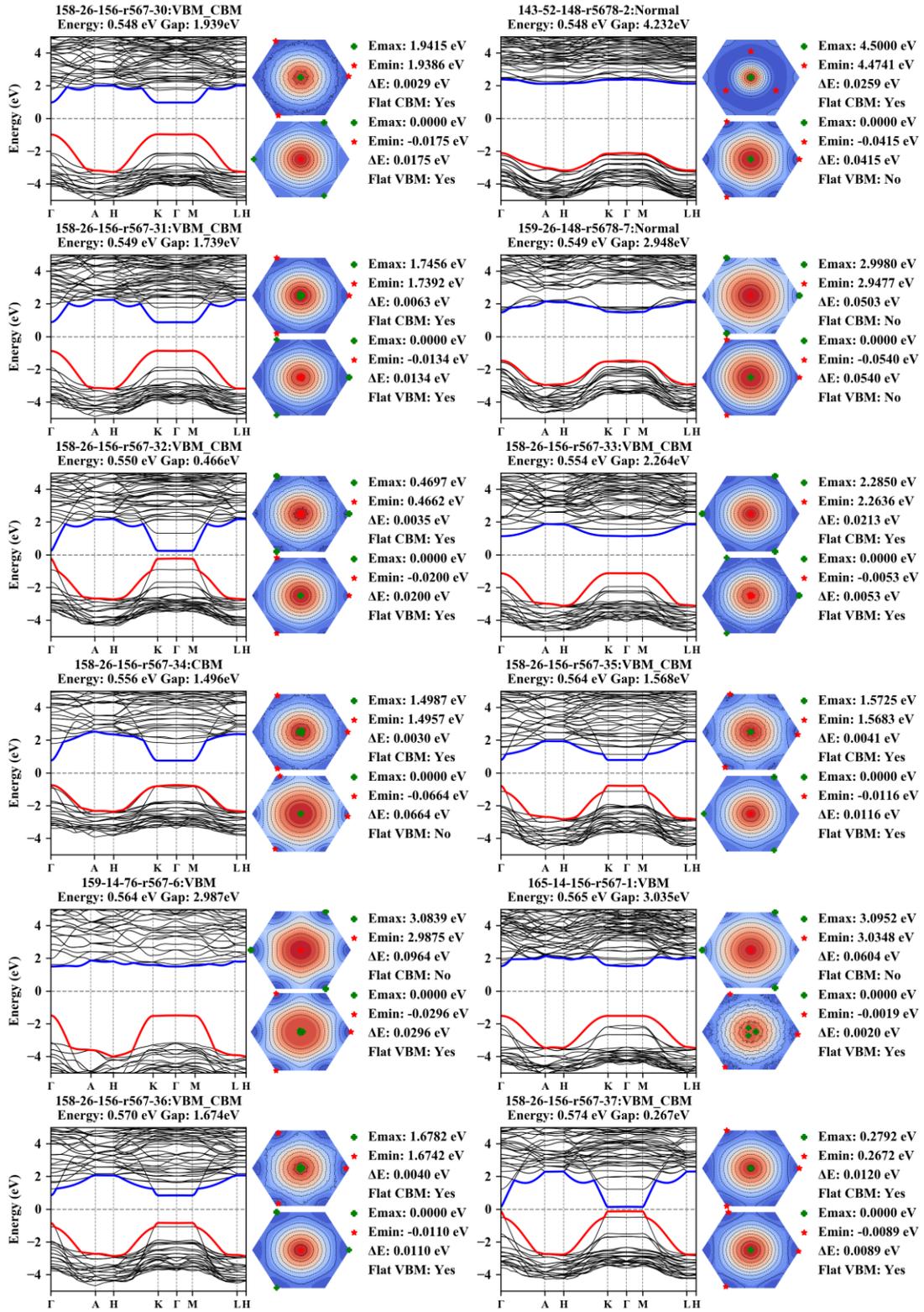

Fig S15. (Structures ranked 133–144 in terms of stability) 2D tight-binding band structures along conventional high-symmetry paths and 3D representations of the highest valence and lowest conduction bands sampled in the Kz=0 plane of the 2D BZ. The total energy, band gap, maxima and minima of valence and conduction bands, band fluctuations in the 2D BZ are provided.

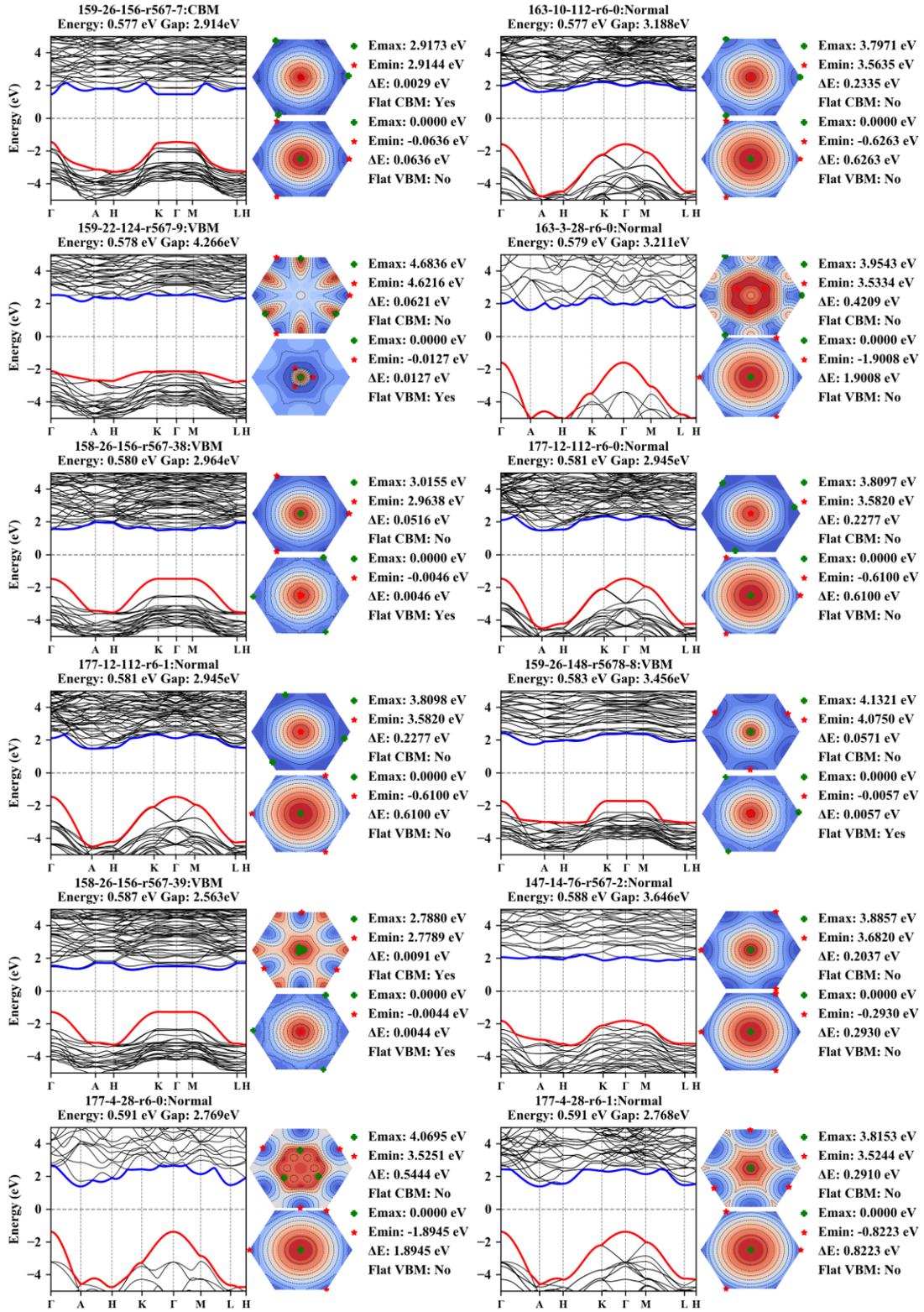

Fig S16. (Structures ranked 145–156 in terms of stability) 2D tight-binding band structures along conventional high-symmetry paths and 3D representations of the highest valence and lowest conduction bands sampled in the Kz=0 plane of the 2D BZ. The total energy, band gap, maxima and minima of valence and conduction bands, band fluctuations in the 2D BZ are provided.

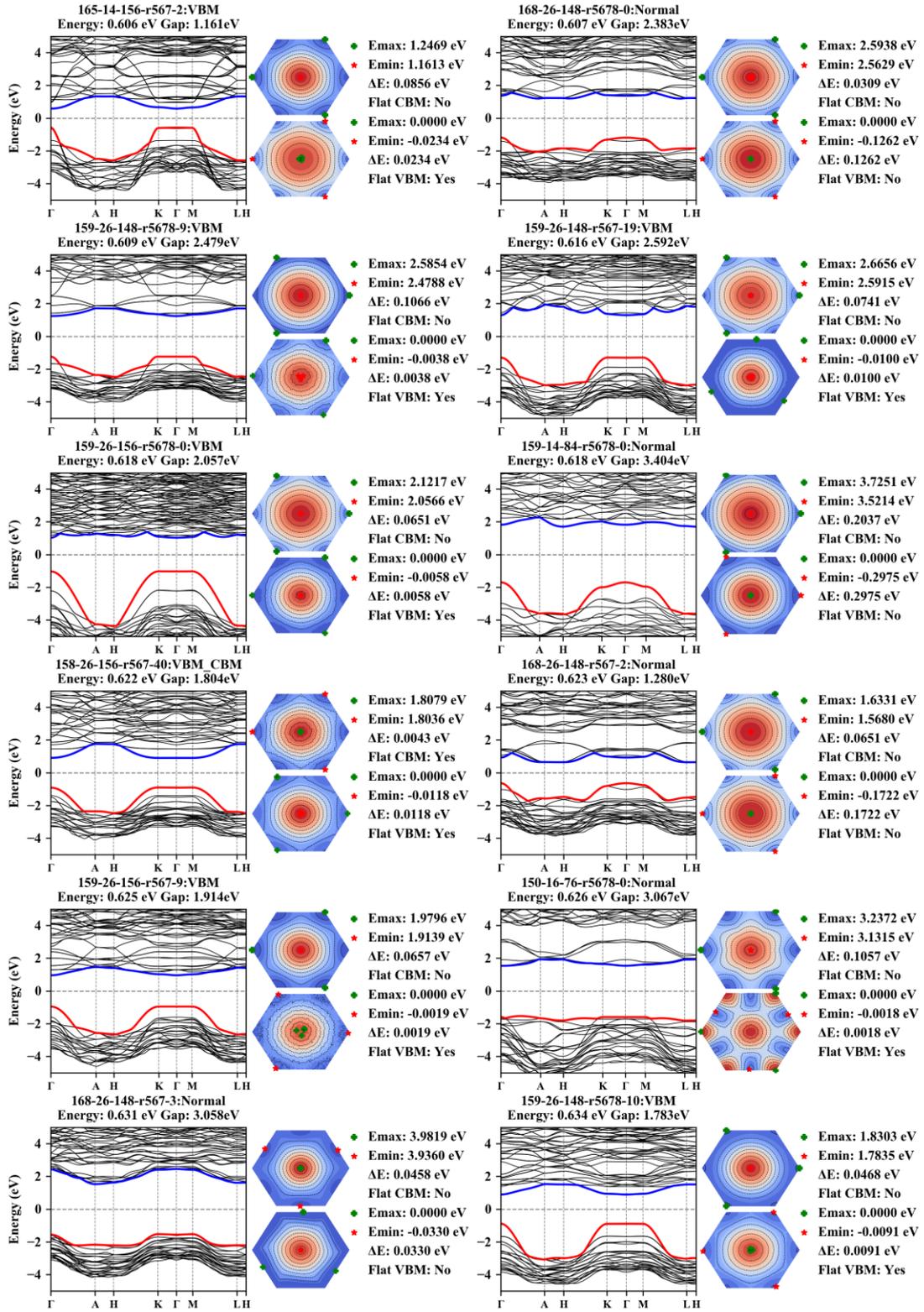

Fig S17. (Structures ranked 157–168 in terms of stability) 2D tight-binding band structures along conventional high-symmetry paths and 3D representations of the highest valence and lowest conduction bands sampled in the Kz=0 plane of the 2D BZ. The total energy, band gap, maxima and minima of valence and conduction bands, band fluctuations in the 2D BZ are provided.

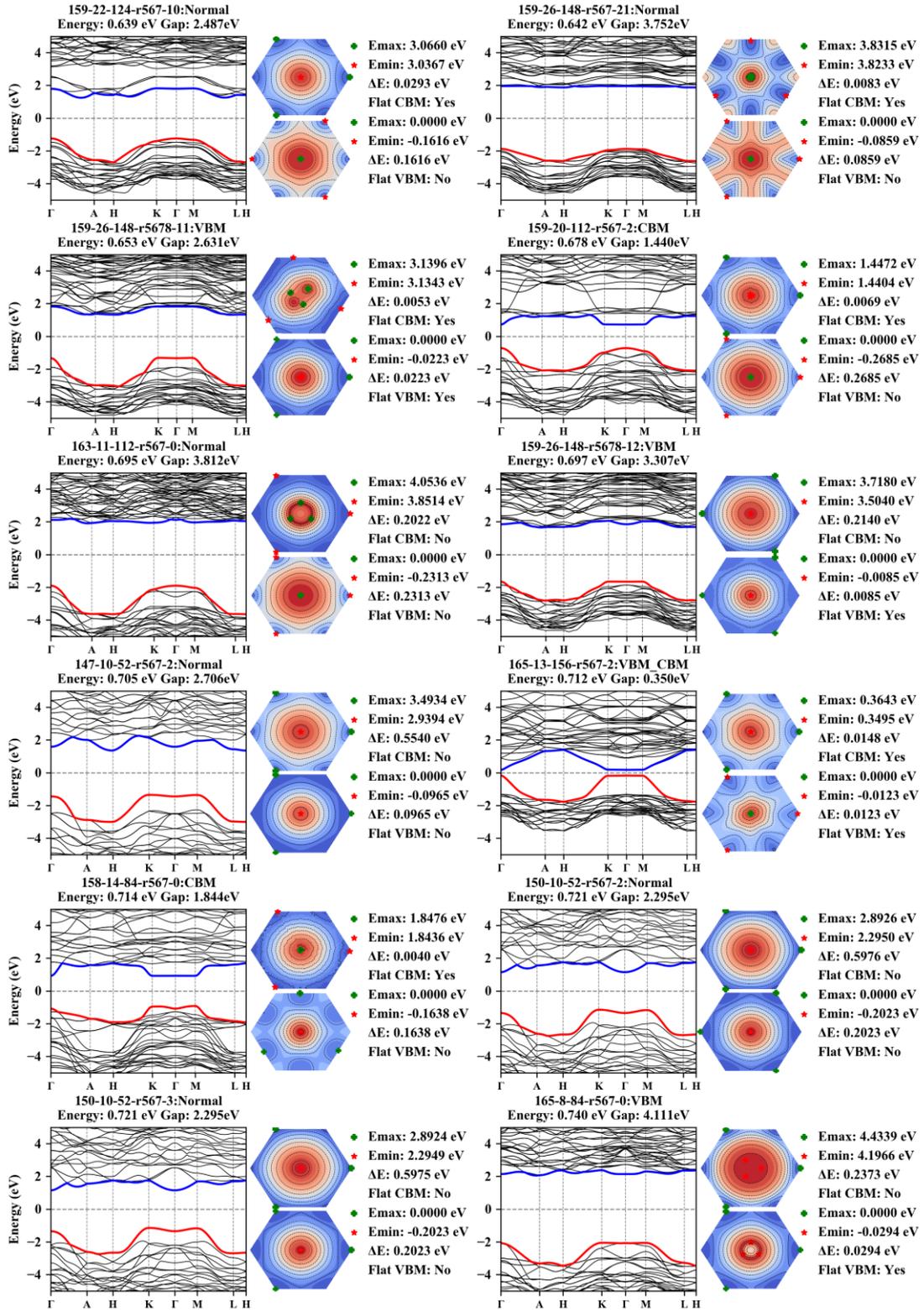

Fig S18. (Structures ranked 169–180 in terms of stability) 2D tight-binding band structures along conventional high-symmetry paths and 3D representations of the highest valence and lowest conduction bands sampled in the Kz=0 plane of the 2D BZ. The total energy, band gap, maxima and minima of valence and conduction bands, band fluctuations in the 2D BZ are provided.